\documentclass[a4paper,fleqn,usenatbib]{mnras}

\usepackage{newtxtext,newtxmath}

\usepackage[T1]{fontenc}
\usepackage{ae,aecompl}
\usepackage{xcolor}

\usepackage{graphicx}	

\title[Continuum elimination for dust studies]{The Effect of Continuum Elimination in Identifying Circumstellar Dust around Mira}
\author[Shepard \& Speck]{
  Lisa M. Shepard$^{1}$\thanks{E-mail: lisashepard@mail.missouri.edu (LS)}
  and 
  Angela K. Speck$^{2}$\thanks{E-mail: angela.speck@utsa.edu (AKS)}\\
$^{1}$Department of Physics \& Astronomy, University of Missouri, Columbia, MO, 65211, USA\\
$^{2}$Department of Physics \& Astronomy, University of Texas, San Antonio, San Antonio, TX 78249, USA
}

\date{Accepted XXX. Received YYY; in original form ZZZ}

\pubyear{2021}

\begin{document}
\label{firstpage}
\pagerange{\pageref{firstpage}--\pageref{lastpage}}
\maketitle

\begin{abstract}
Asymptotic Giant Branch (AGB) stars are major contributors of cosmic dust to the universe. 
Typically, dust around AGB stars is investigated via radiative transfer (RT) modeling, or via simple deconstruction of observed spectra. However, methodologies applied vary. 
Using archival spectroscopic, photometric, and temporal data for the archetypal dusty star, Mira, we identify its circumstellar silicate dust grains. This is achieved by matching the positions and widths of observed spectral features with laboratory data. To do this comparison properly, it is necessary to account for the continuum emission. Here we investigate various ways in which a continuum is eliminated from observational spectra and how it affects the interpretation of spectral features.
We find that while the precise continuum shapes and temperatures do not have a critical impact on the positions and shapes of dust spectral features, it is important to eliminate continua in a specific way.
It is important to understand what contributes to the spectrum in order to remove the continuum in a way that allows comparison with laboratory spectra of candidate dust species.
Our methodologies are applicable to optically thin systems, like that of Mira. Higher optical depths will require RT modeling, which cannot include many different potential astrominerals because there is a lack of complex refractive indices.
Finally, we found that the classic silicate feature exhibited by Mira is not consistent with a real amorphous silicate alone but may be best explained with a small alumina contribution to match the observed FWHM of the $\sim10\mu$m feature.
\end{abstract}




\section{Introduction}
\label{intro}
Asymptotic Giant Branch (AGB) stars are major contributors of cosmic dust to the interstellar medium. Understanding the formation of cosmic dust ejected from these stars is essential to understanding the broader topics of evolution and composition of stellar and interstellar objects in our universe. We identify dust grains in space by matching the positions of observed spectral features with those seen in laboratory spectra.
While radiative transfer modeling should allow us to build a spectrum that
includes all contributions to the observed spectra, we are hampered by a
paucity of appropriate laboratory data for such modeling.
Typically, we need the complex refractive indices of candidate minerals in
order to perform radiative transfer modeling. While there are many simple
absorption or transmission spectra measurements for candidate minerals,
very few minerals have corresponding complex refractive indices. This limits
the application of radiative transfer modeling if we want to determine the
detailed mineralogy of the dust.

In some cases (usually very optically thin scenarios),
we can simply eliminate a continuum contribution to the observed
spectrum to isolate any observed features and measure their basic spectral
parameters (peak and barycentric positions, relative strength, FWHM)
In this paper, we investigate several methods of continuum elimination to verify whether removing the continuum will affect the shape, strength, and position of the spectral features.

In \S~\ref{bg} we discuss relevant background including the theory behind
continuum removal and previous studies that apply a variety of continuum
elimination methods; in \S~\ref{casestudy} we give background information on
Mira ($o$~Ceti) that is relevant to our study; in \S~\ref{measure} we describe
the data and processing used to eliminate the continuum and measure the
spectral features, while in \S~\ref{results} the results of applying these
methods
are described.
\S~\ref{discussion} contains a discussion of the implications of our results. 

\section{background}
\label{bg}
          
\subsection{Fitting the Continuum}
\label{fitting}
To fit and remove the continuum from an observed spectrum, we need to consider
what contributes to the photons we receive.
If the dust shell is not very opaque
(i.e., its optical depth is low), an observed spectrum can be represented by 

\begin{equation}F_{\rm tot}(\lambda) = F_\star(\lambda) + F_{\rm dust}(\lambda)
\label{eqn1}
\end{equation}

Some starlight is always obscured by the dust, which is why this equation is
most applicable to optically thin situations. As optical depth increases, more
starlight is removed and significant radiative transfer occurs, making this
simple equation no longer accurate.

We can then break down the light emitted by the dust, $F_{\rm dust}$ (dustlight)
as follows:

\begin{equation}
 F_\lambda = F_\star + \sum_{i=1,j=1}^{n,m}  C_j \times Q_{\lambda,j} \times B_{\lambda,i}(T_i)
\label{eqn2}
\end{equation}  

\noindent
where each $B_i$ represents a single dust temperature blackbody (of which there are $n$ total), each $Q_{\lambda,j}$ represents the wavelength-dependent extinction efficiency for a single grain type as defined by its size, shape, composition, and crystal structure, and each $C_j$ represents the scale factor for a single grain type (of which there are $m$ total). This means that at each given wavelength the emission from the individual dust species is simply added to give the total emission at that wavelength.

It is possible to subtract the star's contribution to the spectrum by using
either an observed spectrum of a naked star of similar temperature
\citep[like e.g.,][]{Sloan2003A} or by using a stellar atmosphere model
\citep[from e.g.,][]{Allard2016}.
However, the precise details of the stellar spectrum, with all the molecular
absorption features, especially those due to CO and SiO, are sensitive to the
exact temperature and composition of the stellar atmosphere. The goal of this
paper is not to accurately model the underlying star, but to assess the
effects of the processes of deconstructing observed spectra of dusty stars on
extracting data about the dust.
Therefore, rather than applying a detailed model of the stellar atmosphere,
we approximate the star as  a blackbody with a temperature $T_*$.
Subtracting the stellar flux $F_{\star}$ gives:

\begin{equation}
  F_{dust} = \sum_{i=1,j=1}^{n,m} C_j \times Q_j \times B_i
\label{eqn3}
\end{equation}

\noindent
In cases where the dust shell is optically thick, the starlight will be significantly extinguished and simply subtracting the star is difficult.

The spectrum is often simplified to:

\begin{equation}
F_\lambda = C \times Q_\lambda \times B_\lambda(T)
\label{eqn4}
\end{equation}

\noindent
where $B_\lambda(T)$ is the Planck function for a blackbody of temperature $T$,
$Q_\lambda$ is a composite value including contributions from all dust grains
of various sizes, shapes, crystallinities, and compositions, and with a mean
temperature $T$, and $C$ is a
scale factor that depends on the number of dust particles, their geometric
cross section, and the distance to the star. In the literature, this equation
has been applied to the entire spectrum including the star;
or it has been applied after subtraction of a stellar or blackbody spectrum
(see \S~\ref{prev}). Here we determine the sensitivity of the positions  and
shapes of spectral features to the simplification of the spectrum to
Equations~\ref{eqn1}--\ref{eqn4}.

\subsection{Previous studies with continuum elimination}
\label{prev}
Different methodologies have been used to eliminate the continua from stellar
spectra to extract dust spectral features that follow the theory described in
\S~\ref{fitting}.
The usual techniques are continuum subtraction or division, but
the choice of continuum shape also varies. The opacity (optical depth) of the
dust shell should influence which method is used.
As outlined in the previous section, if the circumstellar dust shell is
optically thin, the spectrum of the star should be subtracted from
the total flux to isolate the temperature-dependent spectrum of the dust.
An optically thick dust shell will be dominated by the spectrum of the dust
which will obscure the flux coming from the star. While previous work using
continuum elimination has covered many different types of dust environments, here we focus on previous work on O-rich Asymptotic Giant Branch (AGB) stars.

Previous work which used continuum elimination methods include
\citet{Sylvester1999}, \citet{Dijkstra2005}, and \citet{Speck2008} who all
used the continuum division method. This is equivalent to applying
Equation~\ref{eqn4} to the whole spectrum.
\citet{Sylvester1999} used data from the Infrared Space Observatory (ISO)
Short- and Long-Wavelength Spectrometers (SWS and LWS, respectively) to
examine a sample of seven oxygen-rich AGB stars with dense (optically thick)
circumstellar dust shells. In these OH/IR stars, the dust completely obscures
the stars at visible wavelengths. Mira was also presented for comparison.
Their method used a “pseudo-continuum” spline-fit to approximate the general
shape of this continuum. Division by a spline fitted continuum was also
employed by \citet{Dijkstra2005} who examined 7-14-$\mu$m ISO spectra of a
sample of 12 M-type evolved stars of varying mass-loss rates and therefore
different optical depths.
Both these papers use spline-fitting, which does not necessarily
represent a real, physical cause for the continuum. 
\citet{Speck2008} also applied a
continuum-division method, focused on IRAS 17485-2534, a highly obscured
oxygen-rich AGB star. In this case, the continuum was modeled as a 400K
blackbody rather than a spline fit.
The match of the
observed continuum to a single blackbody temperature would suggest
that we are seeing an isothermal surface within the dust shell.
This represents the depth at which the shell becomes optically
thick. The lack of extra emission at longer wavelengths suggests
that any outlying dust is low enough in density to have an
insignificant contribution to the overall emission \citep[c.f.][]{Speck2009}.

There are many examples of previous studies which used continuum-subtraction
to extract dust spectral features.
\citet{VK1987} examined Infrared Astronomical Satellite (IRAS) Low-Resolution
Spectrometer (LRS) 8-22$\mu$m spectra for a sample of 467 oxygen-rich AGB
stars having a wide variety of optical depths. The continuum was fit with a
power law of the form $F \propto$ $\lambda^\beta$ on either side of the
10-$\mu$m feature. The power-law fit is taken to represent the true continuum
over the feature, and the strength of the feature is measured relative to the
continuum.
\citet{LML1988} used a similar method to study the dust emission seen in
IRAS LRS spectra of 79 MS, S, and SC stars. In this case, they fit a
blackbody energy distribution to both sides of an emission feature and
subtracted this continuum from the observed spectrum.
\citet{LML1990} applied a slightly different continuum to a larger sample of
(291) Mira variables observed with IRAS LRS. In this study, they subtracted a
2500\,K blackbody from each observed spectrum prior to classifying the shape
and position of the $\sim$10$\mu$m spectral feature. 
\citet{SP95} also used IRAS LRS spectra to investigate a sample of variable
oxygen-rich AGB stars and the silicate emission at 10$\mu$m. They isolated
dust emission by modeling the stellar continuum with a modified Planck
function, fitting it to the spectra, and subtracting it. These studies all fit
the star and subtracted it; they just approximated the star differently.

Following on from IRAS, researchers found new details in the spectra of cosmic
dust from the ISO. While examining the 13-$\mu$m dust emission feature in
ISO SWS spectra of optically-thin oxygen-rich circumstellar dust shells,
\citet{Sloan2003A} eliminated the continua by fitting a naked (dust-free)
oxygen-rich AGB star spectrum and subtracting it.
The 131 stars in the sample are mostly AGB with some supergiants and objects
transitioning between AGB and planetary nebula stages.
\citet{Molster2002} used spectra from the ISO SWS and LWS to study a sample of
17 oxygen-rich circumstellar dust shells surrounding evolved stars, ranging
from AGB stars to the planetary nebula phase and including some massive red
supergiants and objects whose status are unclear. They used a spline fit to
approximate the continuum, which they subtracted. 

\citet{Speck2000} analyzed the ground-based (UKIRT) 8$-$13\,$\mu$m spectra of
142 M-type stars, including 80 oxygen-rich AGB stars and 62 red supergiants.
The dust features in the 10-$\mu$m region were examined using normalized,
continuum-subtracted spectra. For each source, a 3000 K blackbody representing
the stellar photosphere was normalized to the spectrum at 8.0-$\mu$m. This was
then subtracted from the observed astronomical spectrum. It is also worth
noting that varying the assumed 3000 K blackbody temperature by +/- 1000 K has
very little effect on the size and shape of the derived continuum-subtracted
dust features, but the precise positions were not measured. This is similar to
the result found by \citet{DePew2006}, where radiative transfer models showed
that the stellar temperature does not have a major effect on positions and
strengths of spectral features. 

These different approaches have also been used to study various types of
objects other than stars.
\citet{MK2007} looked at grain properties in the dust composition in a broad
absorption line quasar observed with the Spitzer Space Telescope and used a
power law to fit the continuum. The method included a continuum-divided
spectrum, assuming the emission region is optically thin in the infrared.
This is to yield the resulting opacity of the dust, and the temperature
dependence has been eliminated by dividing the spectra by the continua.
They also subtracted a spline-fit continuum of each spectrum to provide the
same base line and ran models on the continuum-divided spectrum.
\citet{Peeters2002} used ISO SWS spectra to fit continua for PAH bands using a
sample of 57 sources including reflection nebulae, HII regions, YSOs, evolved
stars, and galaxies. They used a variety of methods to extract the feature
profiles including subtracting a local spine continuum, subtracting a
polynomial of order 1, drawing a general continuum splined through chosen
points, and drawing a continuum under features.  It is worth noting that PAH
spectral features are much narrower than the $\sim$10$\mu$m silicate feature,
and thus, is less affected by difference in the choice of continuum.

Methods for eliminating the continuum have varied, and the effect of different
methodologies has not been adequately explored.
When working with a very optically thick system, the star is hidden visibly,
and one may be able to divide by a blackbody.
This blackbody represents the dust temperature in the circumstellar shell
where the medium transitions from optically thin to optically thick.
Division by the blackbody extracts the dust emission/absorption efficiencies
($Q_\lambda$) as given in Equation~\ref{eqn4}.
When working with an optically thin system, simply subtracting a stellar
continuum ($F_{\star}$) leaves a residual dust spectrum ($F_{\rm dust}$) that still
depends on the dust temperature (see Equation~\ref{eqn3}). This temperature
dependence has been largely neglected when measuring or classifying the residual
dust spectral features.
It should be noted that there are intermediate optical depths, where
significant radiative transfer occurs such that the simplification to either
Equation~\ref{eqn1} (optically thin) or Equation~\ref{eqn4} (optical thick) is
not applicable. In these cases, it is necessary to apply full radiative
transfer modelling.

Below, we explore the effect of applying Equations~1--4 as different pathways to isolating and measuring observed dust features using the optically
thin star, Mira.

\section{Mira as a case study}
\label{casestudy}
Mira has the distinction of being the first known variable star, having been
discovered over 400 years ago\footnote{https://www.star-facts.com/mira/}.
This pulsating AGB star has a period of 332
days, a V-band magnitude ranging 2-10.1, and spectral type M5--9IIIe. The
temperatures corresponding to Mira’s spectral type range over 3420--2667\,K
(see the stellar spectral flux library by \citealt{Pickles1998}).
Mira itself exists as a binary system consisting of the AGB star (Mira A) and
a compact accreting companion (Mira B), first seen optically in 1923
(\citeauthor{Aitken1923}).
The classic $\sim$10$\mu$m silicate spectral feature was first observed in the
late sixties in the infrared (IR) spectra of several M-type giants and
red supergiants (RSGs) including Mira \citep{Gillett68}. 

This well-studied, archetypal, dusty star is used here to demonstrate that
the analysis method affects the interpretation of dust spectral features and
to find which method(s) of continuum elimination should be applied to the
analysis of similar AGB stars.

Many dusty stars have been classified according to their mid-infrared (IR)
spectra obtained by the IRAS LRS. Mira has been classified by \cite{SP98}
as infrared emission class SE8, indicating silicate and oxygen-rich dust
composition with classic narrow silicate emission.
\cite{Kraemer2002} classified Mira as infrared spectral class 2.SEc,
representing strong silicate emission features with peaks at 10--12$\mu$m
and 18--20$\mu$m. \cite{Sloan2003A} extended this work and classified Mira as
IR spectral class 2.SE8, signifying a lack of a 13-$\mu$m feature.
\cite{Speck2000} used UKIRT spectra to classify Mira’s dust features as
silicate AGB A, indicating a “classic” narrow 9.7-$\mu$m silicate feature. 

Mira displays the classic circumstellar $\sim$10\,$\mu$m silicate dust emission
feature seen in many dusty environments. Understanding the nature of
the silicate dust is important to so many astrophysical environments
\citep[e.g.,][]{vk02,draine03,krishna05,krugel08,mann06,casassus01,chiar07,hao05}
but interpreting this classic spectral feature is sensitive to how we extract
it from our observations. In the next section we will investigate this
sensitivity.

\section{Measuring the Spectral Parameters}
\label{measure}

\begin{figure}
\centering
\includegraphics[angle=0,scale=0.6]{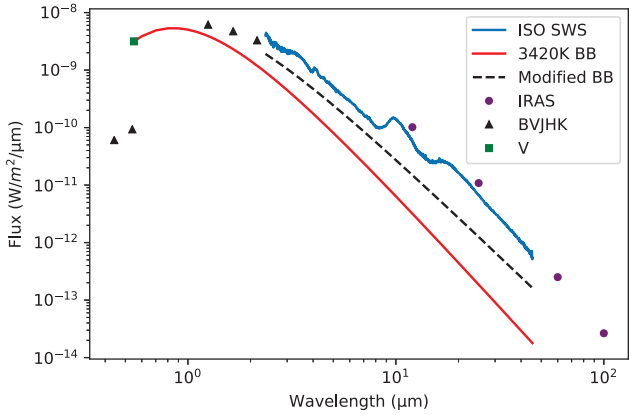}
\caption{The spectral energy distribution for Mira together with the stellar
  blackbody spectrum (T = 3420\,K) and the best fit modified
  blackbody.
  The black triangles are the flux values in the BVJHK bands;
  the purple dots are the flux values in the four IRAS photometry bands;
  the green square is the V-band flux for the same date as the ISO SWS
  observations;
  the blue line is the ISO SWS spectrum of Mira;
  the red line is a 3420\,K blackbody fitted to the observed data at V-band;
  and the black dashed line is the best fit modified blackbody.
  The $x$-axis is wavelength in $\mu$m and
  the $y$-axis is flux density in Wm$^{-2}\mu$m$^-1$.}
\label{MiraSED}
\end{figure} 

\begin{figure*}
\centering
\includegraphics[bb=0 220 13.33in 7.5in,angle=0,scale=0.55]{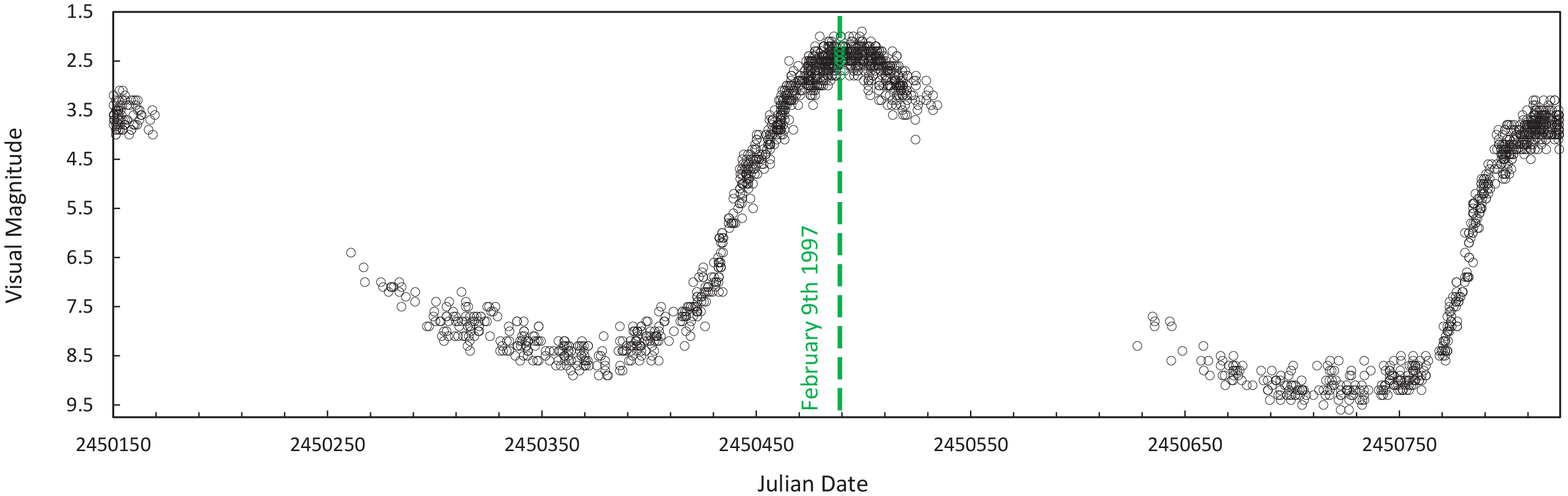}
\caption{The V-band light curve for Mira over a $\sim$22-month period
  (2 full pulsation cycles) from AAVSO. The ISO SWS observation date is
  indicated by the dashed green line. 
 The $x$-axis is Julian Date and the $y$-axis is visual magnitude.}
\label{AAVSO}
\end{figure*}

Mira was observed using ISO SWS on February 9th, 1997. The fully processed post-pipeline spectral data were acquired from an online atlas associated with \citet{Sloan2003B}. Detailed data reduction information is available from the atlas website accessed at https://users.physics.unc.edu/$\sim$gcsloan/library/swsatlas/aot1.html.
We acquired the BVJHK photometric fluxes and the IRAS 12, 25, 60, and 100\,$\mu$m photometry data from the SIMBAD astronomical database\footnote{http://simbad.u-strasbg.fr/simbad/}. In addition to the broad band magnitudes from SIMBAD, we also retrieved the V-band magnitude from the AAVSO\footnote{American Association of Variable Star Observers, https://www.aavso.org/} database for the exact date of the ISO SWS observation.
The resulting spectral energy distribution (SED) is shown in
Figure~\ref{MiraSED}.

The lightcurve from AAVSO shows that the observation date for the ISO SWS
spectrum is close to that of the maximum brightness of Mira
(Figure~\ref{AAVSO}), allowing us to constrain the stellar temperature to the
highest value for known spectral range of  Mira, i.e., 3420\,K.
The V-band photometric data point for the observation date was used to
normalize our stellar (3420\,K) blackbody to the SED as shown
in Figure~\ref{MiraSED}.

\begin{figure*}
\centering
\includegraphics[bb=0 200 13.33in 7.5in,angle=0,scale=0.5]{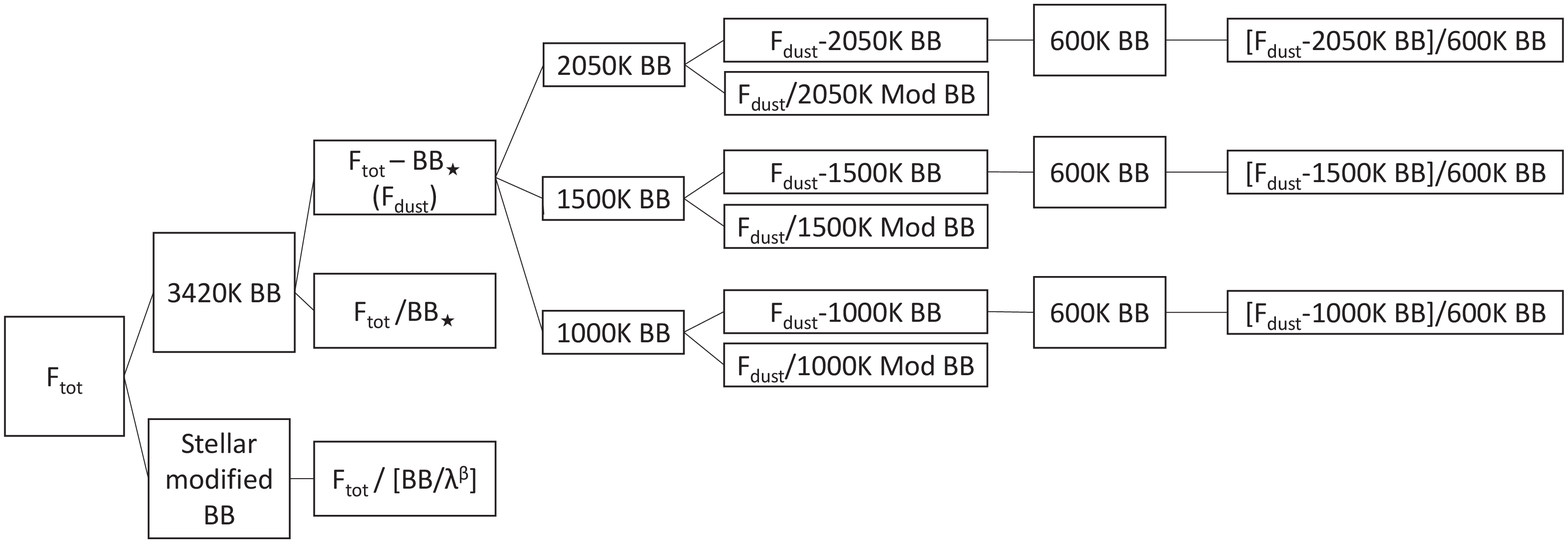}
\caption{Continuum elimination pathways.}
\label{flowchart}
\end{figure*}

\begin{figure}
\centering
\includegraphics[angle=0,scale=0.6]{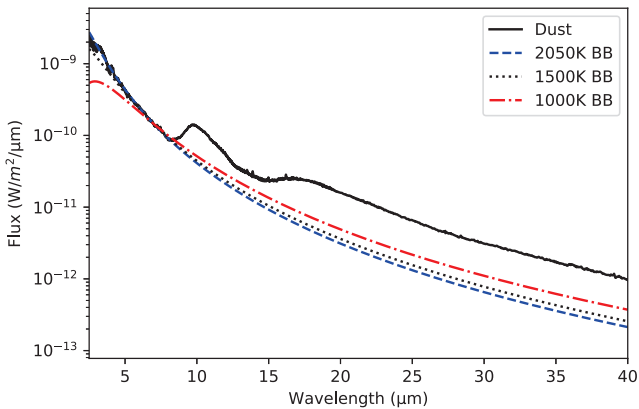}
\caption{F$_{\rm dust}$ - the stellar-blackbody-subtracted, or
  total dust spectrum, for Mira together with the dust continuum blackbodies.
  The black line is F$_{\rm dust}$;
  the dashed blue line is a 2050\,K blackbody;
  the dotted black line is a 1500\,K blackbody; and
  the dot-dash blue line is a 1000\,K blackbody.
  The $x$-axis is the wavelength in $\mu$m
  and the y-axis is flux density in Wm$^{-2}\mu$m$^-1$.}
\label{F_dust}
\end{figure}

To determine the sensitivity of spectral feature measurements to the choice of
continuum elimination, we used six pathways, which fall into two categories: 
1) Fit and subtract a stellar blackbody; or 
2) fold the stellar contribution in equation~\ref{eqn1} into the overall dust flux as in Equation~\ref{eqn4}.

If we subtract a stellar continuum, we must then subtract or divide a dust
continuum. If we subtract a continuum, we must then fit another dust continuum
and divide by it. If we divide by the continuum - we are left with some
form of emission efficiency ($Q$-value).
If we divide by the initial ``stellar'' continuum, we are assuming that
Equation~\ref{eqn4} applies, and we cannot make any more steps.
We can only divide by a continuum once! 
The list of pathways, along with symbols/abbreviations is
summarized in Table~\ref{defn} and a flow diagram to show each pathway is
given in Figure~\ref{flowchart}.

\begin{table}
\caption{Definitions of terms \label{defn}}
\begin{tabular}{lp{5cm}}
\hline
Name/symbol & Definition \\
\hline\hline
F$_{\rm tot}$                          & the total flux coming from Mira and the surrounding circumstellar dust\\
F$_{\rm tot}$/BB$_{\star}$               & total flux received divided by stellar blackbody\\
F$_{\rm tot}$/(BB/$\lambda^{\beta}$)    & total flux received divided by a modified blackbody with emissivity index $\beta$ \\
F$_{\rm dust}$                          & total dust flux = total flux received with stellar blackbody subtracted\\
F$_{\rm dust}$/(BB/$\lambda^{\beta}$)   & total dust flux  divided by a modified blackbody with emissivity index $\beta$ \\
F$_{\rm dust}$-BB$_{\rm dust1}$           &total dust flux with a dust continuum subtracted\\ 
(F$_{\rm dust}$-BB$_{\rm dust1}$)/BB$_{\rm dust2}$ & total dust flux, with a dust continuum subtracted and then divided by a second dust blackbody continuum.\\ 
\hline
\end{tabular}
\end{table}

\begin{figure}
\centering
\includegraphics[angle=0,scale=0.6]{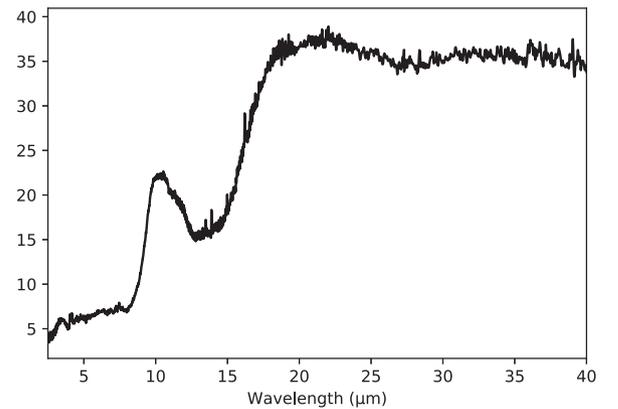}
\caption{The ISO SWS spectrum of Mira divided by 3420\,K stellar
  blackbody. $x$-axis is wavelength in $\mu$m and the
  $y$-axis is normalized flux. }
\label{BBdiv}
\end{figure}

Following Equation~\ref{eqn1}, we can isolate the TOTAL emission from the dust
by subtracting the stellar blackbody. The resulting dustlight spectrum
($F_{\rm dust}$) is shown in Figure~\ref{F_dust}.
Once we subtract the star, the next step is
to fit the dust continuum with either a regular or a modified blackbody where
the modified version can be expressed as $BN = B\lambda^{\beta}$, and $\beta$
is the emissivity index for the dust. The dust continuum is then divided out
to leave a continuum-eliminated spectrum (see e.g., Figure~\ref{BBdiv}). 

We developed a program called MiraFitter which allows us to fit blackbodies
(or modified blackbodies) to any spectrum. The original, flux calibrated
ISO SWS data is $F_{\rm tot} = F_\star + F_{\rm dust}$, the total flux from both
the star and the dust surrounding it. As described in Section~\ref{fitting},
stars with optically thin dust will need the stellar flux subtracted in order
for us to examine the dust.
To fit a blackbody, MiraFitter uses a Planck curve for a given temperature; 
we can select the wavelength at which Planck curve is normalized to the
observational spectral data.
We can also apply a modification via an emissivity law (as described above,
where $\beta$ is the emissivity index).

In addition to fitting blackbodies and/or modified blackbodies, MiraFitter
analyzes the spectral features in each of the continuum-eliminated spectrum
produced by following the 6 pathways shown in Figure~\ref{flowchart} and
Table~\ref{defn} as well as in the original, total flux spectrum (F$_{\rm tot}$).
MiraFitter can find the peak position of the silicate emission features at
$\sim$10\,$\mu$m and $\sim$18\,$\mu$m, calculate the barycentric positions of
those
features, and measure their full width half maxima (FWHM).
While there is often an emphasis on the ``peak position'' for dust spectral
features, both the peak and barycentric positions are important for
discriminating between different potential astrominerals. The combination of
the peak, barycenter, and FWHM gives a good measure of both the position and
the shape of the spectral features. This will be discussed further below and in \S~\ref{discussion}.

To identify the peak positions around $\sim$10\,$\mu$m and $\sim$18\,$\mu$m
in all the various continuum-eliminated spectra, MiraFitter determines the
maximum flux in the ranges 9.5--10.5$\mu$m and 17.6--18.3$\mu$m and then the
corresponding wavelength of each peak.
These positions were examined visually to ensure that the peak measurement is
not due to a noisy point in the observed spectrum. This methodology is applied
to seven different versions of the same original spectrum (as listed in
Table~\ref{defn}).

The barycentric position of a dust spectral feature is often not aligned with
the peak position \citep[see e.g.][]{SWH2011paper}.
While barycenter is often used in terms of the center of mass of a system,
in spectroscopic terms, the barycenter of a spectral feature is the wavelength
at which the energy within the feature is split exactly in half. Therefore, we
need to calculate the area under the curve of the spectral feature and split
that area in half to find the centroid or barycenter.  
The barycentric positions of the $\sim$10$\mu$m and $\sim$18$\mu$m features
as well as the FWHM for the $\sim$10$\mu$m are listed in Table~\ref{peakpos}.
The $\sim$18\,$\mu$m peaks are not consistently defined, and FWHM values are not
reported.
It should be noted that many previous studies of dust features simply fit
the continuum either side of the silicate features at 10 and 18$\mu$m
\citep[e.g.,][]{VK1987,LML1988}, which does yield a clearly defined 18$\mu$m
feature. But the application of this method will shift the position of the
observed feature and make comparison to laboratory data difficult.
This will be discussed further in \S~\ref{discussion}.

As described in Section~\ref{fitting}, Equation~\ref{eqn4} has been applied to
analyzing the whole spectrum including both dust and starlight. 
In this case, the version of the spectrum that is considered continuum
eliminated is F$_{\rm tot}$/BB$_{\star}$,
i.e., the total flux divided by the blackbody spectrum of the star.
This is best applied when the system is optically thick and there is little
{\it direct} starlight to remove \citep[e.g.,][]{Sylvester1999,Speck2009}.
More examples of work that has used this version of continuum elimination are
given in Section~\ref{prev}.

\begin{figure}
\centering
\includegraphics[angle=0,scale=0.6]{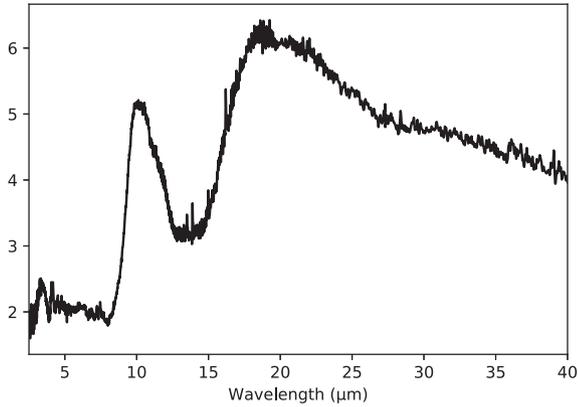}
\caption{The ISO SWS spectrum of Mira divided by a modified blackbody.
  $x$-axis is wavelength in $\mu$m and the
  $y$-axis is normalized flux.}
\label{modBBdiv}
\end{figure}

The F$_{\rm tot}$/BB$_{\star}$ spectrum is shown in
Figure~\ref{BBdiv}, while the F$_{\rm tot}$/(BB/$\lambda^{\beta}$) spectrum (divided by modified blackbody) is shown in Figure~\ref{modBBdiv}.

When the system is optically thin enough that
the total spectrum ($F_{\rm tot}$) is dominated by {\it direct}  starlight,
we must subtract the stellar blackbody to isolate the emission from dust
($F_{\rm dust} = F_{\rm tot}$ – $F_{\star}$, see Equation~\ref{eqn1}). 
There is always some uncertainty in the stellar temperature, especially for variable
stars. In many studies, AGB stars have been assumed to be 3000\,K for simplicity.
Mira itself has a range of temperatures from 2667\,K to 3420\,K.
To determine whether the precise temperature of a stellar continuum affects the
peak position of the spectral dust feature, we varied the stellar blackbody
temperature. 
We found that temperatures above 2493\,K and above could be used before the peak
positions around $\sim$10\,$\mu$m and $\sim$18\,$\mu$m are affected.
The TOTAL dust spectrum (i.e., the stellar-blackbody-subtracted spectrum)
F$_{\rm dust}$ is plotted in Figure~\ref{F_dust}.

\begin{figure}
  \centering
\includegraphics[angle=270,scale=0.3]{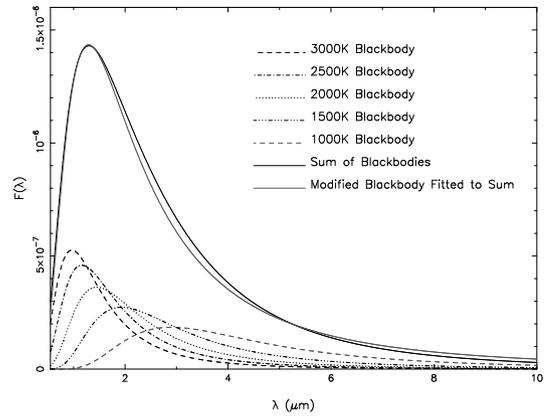}
\caption{A demonstration of the effect of summing blackbodies on the
  continuum-fitting process. $x$-axis is wavelength in $\mu$m; $y$-axis is flux
  density in arbitrary units.}
\label{beta}
\end{figure}

Applying a blackbody modified by an emissivity law (BB/$\lambda^{\beta}$) as the
continuum may be appropriate, depending on the precise nature of the dust.
The emissivity index, $\beta$, depends on the size, structure, composition, and
temperature of the dust grains as well as the optical depth of the dust shell
\citep[e.g.][and references therein]{Demyk2017}.
While in reality $\beta$ should have a value in the range 1-2 to have a
physical meaning \citep{Koike1987,Mennella},
laboratory experiments have shown that it is possible to have emissivity
indices outside this range
\citep[][where amorphous carbon has a $\beta\approx 0.8$]{Koike1980}.
Moreover, a negative $\beta$ value can be used to mimic a continuum that is
the sum of several blackbodies of differing temperatures.
Figure~\ref{beta} demonstrates that using a
negative $\beta$ value emulates the summing of several blackbodies of different
temperatures into a single spectrum. In this case, the value of $\beta$
reflects the range of temperatures and the total amount of dust at each
temperature.

For fitting a modified blackbody to our Mira SED, the value of $\beta$ was
varied until the best fit (by eye) was achieved. This is established using
a visual tool within MiraFitter that displays both the spectrum being fitting
and the fitting modified blackbody as well as a plot of the resultant
divided spectrum. The presence of both the silicate features and the numerous
molecular absorption feature makes a chi-squared minimization difficult to
apply, but the divided spectrum should be flattened
(on average, ignoring the dust and molecular features). 
This fitting process yielded $\beta$ = -0.5.
This best-fitting modified blackbody is shown with the SED in
Figure~\ref{MiraSED}.
A range of $\beta$ values were tested to determine at what values
the peak positions were affected.
The $\sim$10\,$\mu$m peak is unchanged for -0.79 $\leq\beta\leq$ -0.03, and
the $\sim$18\,$\mu$m peak is unchanged for -1.99 $\leq\beta\leq$ 0.0.
It is notable that the beta value is always negative, indicative of a range
of dust temperatures combining to give the continuum (see Figure~\ref{beta}.)
In fact, that we can simply sum a series of blackbodies to replicate a
negative $\beta$ value is consistent with Equation~\ref{eqn3}
where the dust spectrum is a sum of blackbodies with emission efficiencies
applied.

Once we have removed the stellar contribution from the spectrum (i.e., generated
$F_{\rm dust}$), we then fit another blackbody to account for the
temperature-dependent dust continuum.
There are several ways to conceptualize the dust ``continuum''.
It could be the blackbody ($B_\lambda(T)$) from Equation~\ref{eqn4}, in
which case, division by said blackbody would yield the emission efficiency factor
$Q_\lambda$ for a composite dust species including contributions from all dust
grains of various sizes and assuming a single mean temperature.
Alternatively, the continuum could be due to a collection of dust grains that
contribute to the overall continuum but do not display diagnostic spectral features
(e.g., metallic iron). In this case, the continuum would be one of the
$C_j \times Q_j \times B_i$ components in Equation~\ref{eqn3}, and thus
subtracting the continuum to isolate the dust with spectral features is appropriate.
Note that in this case we are still left with another $ C_j \times Q_j \times B_i$
component and this residual spectrum is still temperature dependent.

To test the effect of different dust continua,
we created blackbodies of 1000\,K, 1500\,K, and 2050\,K for removal. The fits of these continuum blackbodies to the total dust spectrum is shown in
Figure~\ref{F_dust}.
The conventional (most common) choice for the dust
continuum temperature is based on the identification of the $\sim$10\,$\mu$m
feature as amorphous or glassy silicate. Approximately 1000\,K is the typical
silicate glass transition temperature \citep[see][]{SWH2011paper}. Above the glass
transition, the silicate grains will either form directly as crystalline
material or an amorphous solid will begin crystallizing.
The other temperatures we chose coincide with
(1) the condensation temperature for olivine and pyroxene; $\sim$1500\,K; and
(2) best fit to the total dust spectrum, F$_{\rm dust}$ without any
theoretical considerations about dust stability; 2050\,K.
While 2050\,K is above the stability temperature for any silicates, it is not
an unfeasible temperature for the first solids to form \citep[e.g.,
  high temperature oxides, aluminates or titanates;][]{McSween1989,Ebel2006}.
Furthermore, there is a molecular layer in the atmosphere of Mira with a
temperature of 1500--2100\,K \citep{Perrin2004}.
In addition, this 2050\,K blackbody was the best fit
for the total dust spectrum and thus serves to test
(a) the sensitivity to getting the best fitting continuum rather than using a
physically expected value and
(b) the effect of temperature on the resulting spectral feature parameters.

Assuming that most of the dust is silicate and that the dust emission
can be modeled according to Equation~\ref{eqn4}, we need to divide by the
dust-continuum temperature to extract the temperature-independent parameters
($Q$-values) of the dust spectral features.

As with the measurement of the spectral feature parameters in the total dust
spectrum (F$_{\rm dust}$), we wanted to test for the effect of varying the
{\em star} temperature used to generate $F_{\rm dust}$ prior to dividing out the
dust continuum: i.e., we want to know whether there are downstream effects
on the spectral feature parameters that depend on early steps in the
pathway to continuum elimination.
After dividing out the dust continuum, the peak position of silicate spectral
features at $\sim$10\,$\mu$m and $\sim$18\,$\mu$m remain unchanged for all 
stellar temperatures greater than 2493\,K, well within the known temperature of
Mira for the observation date.

\begin{figure*}
\centering
\includegraphics[angle=0,scale=1.0]{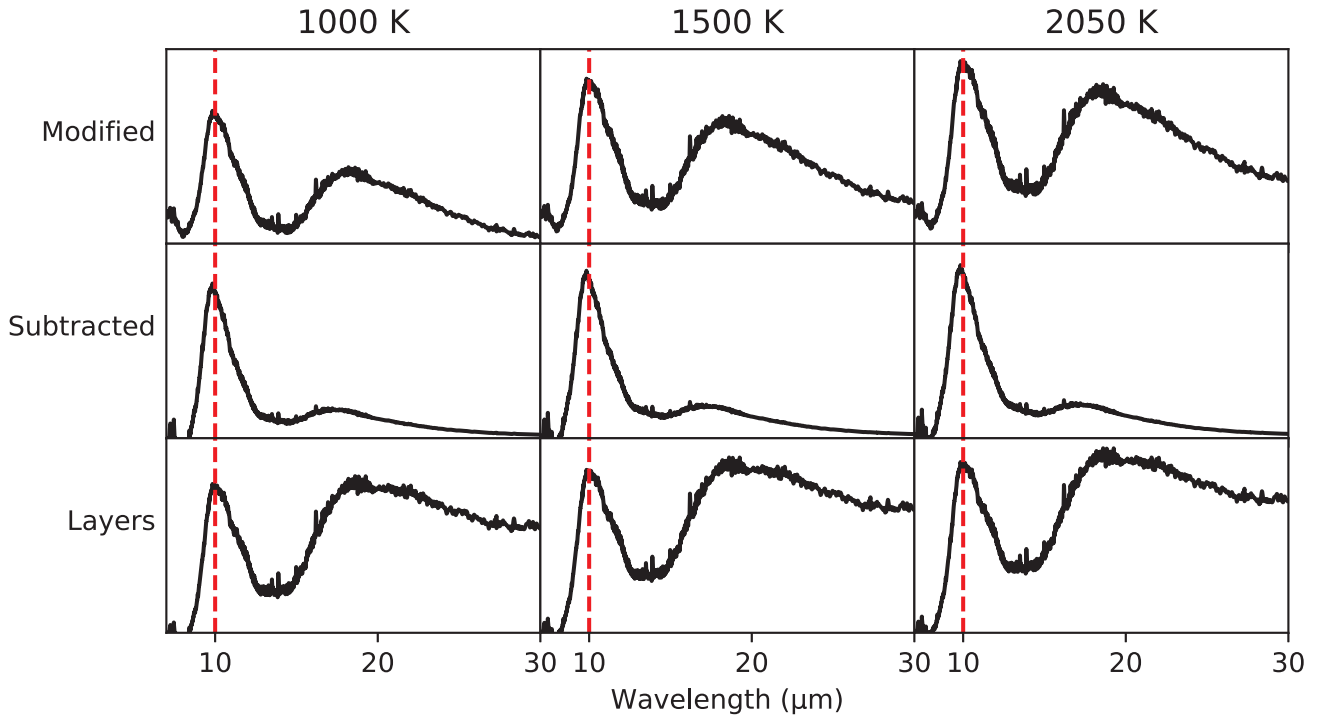}
\caption{The effect of different dust continuum elimination.
  Each column shows the temperature labeled at the top for the blackbody used
  for continuum removal.
  The top row uses a modified blackbody and is the total dust spectrum
  {\it divided} by the blackbody dust continuum for that column,
  i.e.,  F$_{\rm dust}$/(BB/$\lambda^{\beta}$);
  the middle row shows the total dust spectrum with the dust continuum blackbody {\it subtracted}, i.e. F$_{\rm dust}$-BB$_{\rm dust1}$;
  and the bottom row is the same as the middle row, but with a 600\,K blackbody  divided out. i.e., (F$_{\rm dust}$-BB$_{\rm dust1}$)/BB$_{\rm dust2}$.}
\label{array}
\end{figure*}

In addition, F$_{\rm dust}$/(BB/$\lambda^{\beta}$) uses emissivity after
subtracting the star. A new blackbody is constructed using dust temperatures
of 1000, 1500 and 2050\,K, as described above, normalized at 7 \,$\mu$m.
Each blackbody was modified by an emissivity law;
the emissivity index, $\beta$, was varied until the best fit
(by eye, using the methodology described above) was
achieved. This gave $\beta$ = -0.9, which works for all temperatures above
411\,K. The resulting spectra are shown in Figure~\ref{array} (top row).
A range of $\beta$ values was tested to determine its effect on spectral
feature parameters. The peak positions remain unchanged
for -0.74 $\leq$ $\beta$ $\leq$ -1.96. The position and shape
of the spectral features is not very sensitive to the temperature or value of
emissivity index.

It is possible that the blackbody fitted to the star-subtracted spectrum
(F$_{\rm dust}$) represents a different dust component than silicate (for example,
metallic iron or some form of high-temperature oxide). In that case, it is
necessary to subtract the dust blackbody to isolate the emission from silicate.
Then we need to fit a new blackbody to account for the temperature of the silicate
dust. Figure~\ref{array} (middle row) shows this F$_{\rm dust}$-BB$_{\rm dust1}$,
with the dust blackbody curve subtracted from the original F$_{\rm dust}$.
The spectral features at $\sim$10\,$\mu$m and $\sim$18\,$\mu$m in the F$_{\rm dust}$-BB$_{\rm dust1}$ (dust continuum subtracted spectrum) have peak positions
that remain the same for a dust temperature range of 1000--3200\,K.

The final step is to determine the intrinsic spectral parameters for the
silicate features in this 2-dust component scenario.
Since this leaves a residual spectrum that is still temperature-dependent,
we then fit a second blackbody curve for a dust layer at 600\,K, which was
divided out to give us the temperature-independent $Q$-values, i.e.,
(F$_{\rm dust}$-BB$_{\rm dust1}$)/BB$_{\rm dust2}$, shown in
Figure~\ref{array} (bottom row).
The $\sim$10\,$\mu$m peak is unchanged for the second dust layer having
temperatures ranging 203$-$663\,K, and the $\sim$18\,$\mu$m peak is
unchanged for temperatures above 225\,K.

\section{Measurement Results}
\label{results}

The results of the spectral feature measurements described in \S~\ref{measure}
are summarized in Table~\ref{peakpos}.
We have tested the effect on spectral parameters of using different
temperatures and emissivity indices and found that the measurements
are not very sensitive to either temperature or emissivity index. However,
Table~\ref{peakpos} shows that the precise pathway to continuum elimination
can have a large impact on the peak position, barycenter, and FWHM of the
spectral features.

For the $\sim$10\,$\mu$m feature, the peak position is consistent for all pathways
to continuum eliminations that start with subtraction of the stellar
blackbody, whereas the $\sim$18\,$\mu$m is more sensitive to the next steps in
the continuum elimination.
For the pathways to continuum elimination that do not start with subtracting the star, even the $\sim$10\,$\mu$m feature varies.
The barycentric positions for both the $\sim$10\,$\mu$m and $\sim$18\,$\mu$m features
shows similar trends with the $\sim$18\,$\mu$m peak position, showing that
division by a continuum has a major effect on how we measure the position of
the silicate features.
The implications of these results are discussed in \S~\ref{discussion}.

\begin{table*}
\caption{Spectral Feature Measurements \label{peakpos}}
\begin{tabular}{lrrrrr}
\hline
& \multicolumn{2}{c}{Peak position ($\mu$m)}&\multicolumn{2}{c}{Barycenter}& FWHM\\
& $\sim10\mu$m &  $\sim18\mu$m & $\sim10\mu$m &  $\sim18\mu$m & ($\mu$m)\\
\hline\hline
F$_{\rm tot}$                          & 9.83 & 17.64 & 10.15 & 17.75 & 2.18\\
F$_{\rm tot}$/BB$_{\star}$                & 10.37 & 18.20 & 10.49 & 19.06 & 2.87\\
F$_{\rm tot}$/(BB/$\lambda^{\beta}$)    & 10.12 & 18.20 & 10.44 & 18.97 & 2.77\\
F$_{\rm dust}$                          & 9.83 & 17.64 & 10.16 & 17.74 & 2.05\\
1000\,K F$_{\rm dust}$/(BB/$\lambda^{\beta}$)   & 9.83  & 18.20 & 10.36 & 18.83 & 2.36\\
1500\,K F$_{\rm dust}$/(BB/$\lambda^{\beta}$)   & 9.83  & 18.20 & 10.39 & 18.86 & 2.47\\
2050\,K F$_{\rm dust}$/(BB/$\lambda^{\beta}$)   & 9.83  & 18.20 & 10.40 & 18.88 & 2.54\\
1000\,K F$_{\rm dust}$-BB$_{\rm dust1}$              & 9.83  & 17.64 & 10.20 & 18.27 & 2.36\\
1500\,K F$_{\rm dust}$-BB$_{\rm dust1}$              & 9.83  & 17.64 & 10.19 & 18.20 & 2.34\\
2050\,K F$_{\rm dust}$-BB$_{\rm dust1}$              & 9.83  & 17.64 & 10.19 & 18.16 & 2.34\\
1000\,K (F$_{\rm dust}$-BB$_{\rm dust1}$)/BB$_{\rm dust2}$ & 9.83 & 18.20 & 10.37 & 18.93 & 2.70\\
1500\,K (F$_{\rm dust}$-BB$_{\rm dust1}$)/BB$_{\rm dust2}$ & 9.83 & 18.20 & 10.37 & 18.93 & 2.76\\
2050\,K (F$_{\rm dust}$-BB$_{\rm dust1}$)/BB$_{\rm dust2}$ & 9.83 & 18.20 & 10.37 & 18.93 & 2.76\\
\hline
\end{tabular}
\end{table*}

\section{Discussion}
\label{discussion}

\begin{figure}
\centering
\includegraphics[angle=270,scale=0.55]{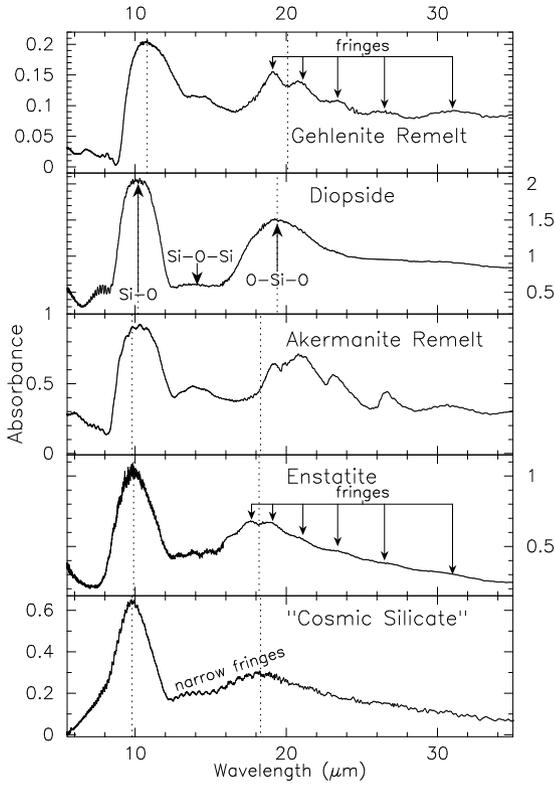}
\caption{Laboratory absorbance spectra for samples from \citep{SWH2011paper}.
  In all cases the $x$-axis is the wavelength in $\mu$m and the $y$-axis is
  absorbance. The dotted line indicates the barycentric positions for the
  $\sim$10\,$\mu$m and $\sim$18\,$\mu$m features. The wavelengths of the peak
  positions are listed in Table~\ref{tab:glasses}.}
\label{glasses}
\end{figure}

We investigated the positions and shapes of the
$\sim$10\,$\mu$m and $\sim$18\,$\mu$m spectral features using different
analysis methods.

\begin{figure}
\centering
\includegraphics[angle=270,scale=0.6]{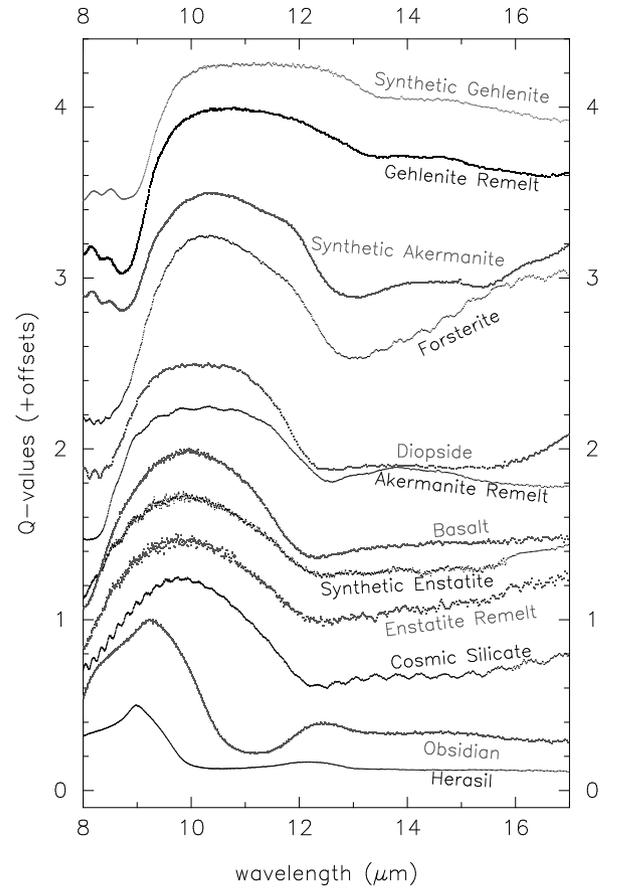}
\caption{Calculated Q-values (absorption efficiency factors) for samples
  from \citet{SWH2011paper}.
  These Q-value spectra are stacked to demonstrate the
  shift in wavelength of the peak of the $\sim$10\,$\mu$m with composition.}
\label{SWH2011}
\end{figure}

We identify dust components in circumstellar spectra by comparing with
laboratory data. The precise composition and structure of a silicate
will affect the parameters of the $\sim$10\,$\mu$m and $\sim$18\,$\mu$m of
spectral features. 
Comparing the laboratory data parameters from \citet{SWH2011paper},
shown in Table~\ref{tab:glasses} and Figures~\ref{glasses} and \ref{SWH2011},
and the peak positions and barycenters extracted from the Mira spectrum
(shown in Table~\ref{peakpos}), we can see that incorrect continuum
 elimination will lead to incorrect attributions of minerals to
 circumstellar features.

\begin{table}
 \centering
 \caption{Spectral Parameters of Glasses from \citet{SWH2011paper}
   (all units are $\mu$m)
   \label{tab:glasses}}
\begin{tabular}{lrrrrr}
\hline
Sample & \multicolumn{2}{c}{Peak } &  \multicolumn{2}{c}{Barycenter}& FWHM  \\
Name   & $\sim$10$\mu$m  & $\sim$18$\mu$m & $\sim$10$\mu$m  & $\sim$18$\mu$m  & $\sim$10$\mu$m \\
\hline\hline
Gehlenite	&	10.3	&	       	&	10.8	&		&	2.86	\\
Akermanite      &	10.3	&		&	10.6	&		&	2.46	\\
Forsterite	&	10.2	&	       	&	10.4	&		&	2.43	\\
Diopside       	&	10.1	&	19.2	&	10.2	&	19.4	&	2.46	\\
Basalt         	&	10.0	&	       	&	9.9	&		&	2.41	\\
Enstatite	&	9.9	&	17.6	&	9.9	&	18.5	&	2.58	\\
Cosmic silicate	&	9.8	&	18.3	&	9.7	&	18.7	&	3.15\\
Obsidian	&	9.0	&	       	&	9.1	&		&	2.14	\\
Herasil (SiO$_2$)&	9.0	&		&	8.9	&		&	1.34	\\
\hline
\end{tabular}
\end{table}


For instance, the $\sim$10\,$\mu$m feature peaks at 9.8-9.9$\mu$m for both
Enstatite and ``Cosmic Silicate''%
\footnote{\citet{SWH2011paper} produced infrared spectra of a silicate glass
  for which the ratios of the major cations (Mg, Si, Al, Na, Ti) were the
  same as those of chondrites/the solar system at large but excluded iron.
  This glass sample was then used to produce the complex refractive index for
  Cosmic Silicate, where the same sample was measure spectroscopically from
  0.2$\mu$m to 200$\mu$m \citep{SPH2015}.}
but their
$\sim$18\,$\mu$m features peak at 17.6$\mu$m and 18.3$\mu$m, respectively.
This means that F$_{\rm tot}$ and F$_{\rm dust}$ are consistent with Enstatite, but
F$_{\rm dust}$/(BB/$\lambda^{-\beta}$) and
(F$_{\rm dust}$-BB$_{\rm dust1}$)/BB$_{\rm dust2}$ are consistent
with ``Cosmic Silicate''. Meanwhile the continuum-elimination pathways
that do not subtract the star would suggest a match to Forsterite or Gehlenite.

When we focus on the barycentric position of the features
(rather than peak position),
we see similar trends with the $\sim$18\,$\mu$m peak position.
The barycentric position of both silicate features is shifted whenever a
continuum division is involved in the process of continuum elimination.

Comparing the measured barycenters to those for laboratory samples listed in
Table~\ref{tab:glasses}, we see that the  $\sim$10\,$\mu$m feature matches
diopside (CaMgSiO$_{\rm 4}$) most closely when there is no continuum division in
the elimination pathway but is closer to forsterite for continuum elimination that uses a divided continuum.
Unfortunately, we do not have a measurement of the
$\sim$18\,$\mu$m feature for this forsterite glass, but the diopside 
$\sim$18\,$\mu$m feature occurs at a significantly longer wavelength than
that observed for Mira in any version of the continuum elimination.

The comparison of both the peak position AND the barycentric position of the
dust spectral features is critical to identifying their mineral carriers.
Simply measuring the peak position will not suffice.

In addition to the peak and barycentric positions of the observed spectral
features, we also measured the FWHM of the $\sim$10\,$\mu$m feature.
It is clear from the list in Table~\ref{peakpos} that the FWHM is quite
sensitive to the choice of continuum elimination pathway. Unlike the peak
position and barycenter, the FWHM is sensitive to the choice of continuum
temperature, making the parameter difficult to use to identify potential
mineral carriers.

We need to compare like with like, i.e., ensure that what we take from
laboratory data is equivalent to observational data to which we compare.
This is discussed extensively in \citet{Speck2013}. 
Figure~\ref{SWH2011} shows the $Q_\lambda$ values for a range of silicate glass
compositions, while Figure~\ref{glasses} shows the wavelength dependent
absorbance for the silicate glasses for which \citet{SWH2011paper} measured the 
$\sim$18\,$\mu$m feature. Absorbance is closer to an optical depth than a
$Q$-value.

Looking back at \S~\ref{fitting}, we can see for a
fairly optically thin dust shell like that of Mira%
%
, we should subtract the
stellar continuum (see Eqn~\ref{eqn2}). This leaves us with a temperature-
dependent dust spectrum as given in Eqn~\ref{eqn4}.
To extract the emission/absorption efficiency, we must divide the observed
spectrum by the blackbody temperature of the dust. Without dividing by the
dust continuum AFTER subtraction of the star, we should not compare to the
$Q$-values.

Many of the studies of silicate features in circumstellar dusts, like those
mentioned in \S~\ref{prev}, simply subtract the stellar continuum and do
not completely eliminate the dust continuum. Thus, those spectra cannot be
used to infer the mineralogy of the dust.

If we follow any of the continuum elimination pathways that use both stellar
continuum subtraction and dust continuum division, we find that for Mira the
peak positions of the silicate features are consistently at
9.83 and 18.20$\mu$m, while the barycenters occur at 10.4 and 18.9$\mu$m
(with a little more variability than the peak position).
The peak position is consistent with ``Cosmic Silicate'' while the
barycentric position is not. This may suggest the need for a non-silicate
component \citep[e.g., alumina;][]{Speck2000},
even in this archetypal strong silicate feature.

It is clear from this study that care must be taken when deconstructing
observational spectroscopic data. While radiative transfer modeling may be
able to provide a more accurate representation of the temperatures of dust
involved in the emission of photons at each wavelength, RT modeling is
hampered by a lack of applicable laboratory data. Most radiative transfer
modeling uses synthetic complex refractive indices (or dielectric constants)
such as \citet{dl1985,ohm1992}. However, these optical constants are not based
on real mineral samples and cannot allow us to extract mineralogical
information from observed spectra
\citep[see][and references therein]{SPH2015}. Moreover, there are problems
with many of the published refractive indices that are based on real mineral
samples \citep{SWH2011paper}. Consequently, it is difficult to use RT modeling
when the goal is to extract the detailed dust mineralogy.

Here we have shown that for low optical depth systems, we MUST subtract a
stellar contribution from the observed total spectrum
(see Equation~\ref{eqn1}) - but the precise $T_\star$ value is not critical.
Then we must divide by a dust continuum in
order to get a spectrum equivalent to the emission efficiency $Q_\lambda$
(i.e., applying Equation~\ref{eqn4}).
In this case we generate F$_{\rm dust}$/(BB/$\lambda^{\beta}$) and include the
step of subtracting a featureless continuum due to e.g., metallic iron will
not change the relevant parameters of the silicate features
(although the overall feature-to-continuum ratio may be affected).
Failure to subtract a stellar continuum OR divide by a dust continuum will
lead to incorrect feature parameters and thus to misidentified astrominerals.
As optical depth increases, the application of this simplified deconstruction
method becomes problematic and RT modeling becomes necessary.
However, at very high optical depth, there are several examples of systems
where the continuum can be fitted by a single blackbody
\citep[e.g.,][]{Speck2008,Speck2009}. In this case we are
observing an isothermal layer in the dust shell outside of which the dust is
optically thin. We usually apply Equation~\ref{eqn4} to determine the
$Q_\lambda$  values for the absorbing dust in the optically thin outer layers,
but further study should be undertaken to determine what affect this has,
versus treating the isothermal layer as the central ``stellar'' source.

\section{Conclusions}
We find that for optically thin dust shells, it is possible to deconstruct
their spectra to extract detailed mineralogical information. However, the
precise pathway by which continua are eliminated has
significant effects on the residual dust spectrum and may lead to
misidentification of dust species in space.
For moderate to high optical depth systems, we cannot apply this simple
deconvolution and must apply RT modeling. Unfortunately, RT modeling is
hampered by a paucity of complex refractive indices for detailed
mineralogical studies. To continue investigating dust in space we need
to (a) acquire optical constants (complex refractive indices) of more real
mineral samples; and (b) revisit the many studies of optically thin systems
that have classified the silicate features with a new eye to understanding the
detailed mineralogy.

Finally, our study suggests that the archetypal/classic silicate feature
exhibited by Mira is not consistent with a real amorphous silicate alone, but
may be best explained with a small alumina contribution to match the observed
FWHM of the $\sim10\mu$m feature.


\section*{Acknowledgements}
The authors are grateful to the AAS for the opportunities afforded to present
preliminary versions of this work at their meetings, and especially for the
Chambliss Medal awarded to LS for her poster on this topic
(AAS Winter meeting \#235 in Hawaii in 2020).


\section*{Data Availability}
The data underlying this article are available in publicly.
The datasets were derived from sources in the public domain:
(1) ISO SWS spectroscopic observational data from
https://users.physics.unc.edu/$\sim$gcsloan/library/swsatlas/aot1.html;
(2) all photometric data points were retrieved from SIMBAD
(http://simbad.u-strasbg.fr/simbad/); and
(3) lightcurve data from the
AAVSO (https://www.aavso.org/).




%



\bsp	
\label{lastpage}
\end{document}